\documentclass[aps,twocolumn] {revtex4-1}
\usepackage{amsmath,graphicx,amsfonts,amssymb,bbm,times}
\usepackage[colorlinks=true,linkcolor=blue,urlcolor=blue,citecolor=blue, breaklinks=true,hypertexnames=false]{hyperref}
\usepackage[utf8]{inputenc}
\DeclareUnicodeCharacter{03BC}{\mbox{$\mu$}}
\DeclareUnicodeCharacter{03B4}{$\delta$}
\DeclareUnicodeCharacter{2009}{-}

\usepackage[english]{babel}
\addto\captionsenglish{}
\addto\captionsenglish{}

\makeatletter
\def\bbl@set@language#1{%
	\edef\languagename{%
		\ifnum\escapechar=\expandafter`\string#1\@empty
		\else\string#1\@empty\fi}%
	\@ifundefined{babel@language@alias@\languagename}{}{%
		\edef\languagename{\@nameuse{babel@language@alias@\languagename}}%
	}%
	\select@language{\languagename}%
	\expandafter\ifx\csname date\languagename\endcsname\relax\else
	\if@filesw
	\protected@write\@auxout{}{\string\select@language{\languagename}}%
	\bbl@for\bbl@tempa\BabelContentsFiles{%
		\addtocontents{\bbl@tempa}{\xstring\select@language{\languagename}}}%
	\bbl@usehooks{write}{}%
	\fi
	\fi}
\newcommand{\DeclareLanguageAlias}[2]{%
	\global\@namedef{babel@language@alias@#1}{#2}%
}
\makeatother

\DeclareLanguageAlias{en}{english}

\makeatletter
\def\@bibdataout@aps{%
	\immediate\write\@bibdataout{%
		@CONTROL{%
			apsrev41Control%
			\longbibliography@sw{%
				,author="08",editor="1",pages="1",title="0",year="1"%
			}{%
				,author="08",editor="1",pages="1",title="",year="1"%
			}%
		}%
	}%
	\if@filesw \immediate \write \@auxout {\string \citation {apsrev41Control}}\fi 
}
\makeatother

\def\vep{\varepsilon}
\def\wt{\widetilde} 

\def\da{\dagger}

\def\dd{\mathrm d}

\def\da{\dagger}

\date\today
\begin{document}
	
	\title{Dissipative dynamics of a heavy impurity in a Bose gas in the strong coupling regime}
	\author{Aleksandra Petkovi\'{c} and Zoran Ristivojevic }
	\affiliation{Laboratoire de Physique Th\'{e}orique, Universit\'{e} de Toulouse, CNRS, UPS, 31062 Toulouse, France}

	\begin{abstract}
		We study the motion of a heavy impurity in a one-dimensional Bose gas. The impurity experiences the friction force due to scattering off thermally excited quasiparticles. We present detailed analysis of an arbitrarily strong impurity-boson coupling in a wide range of temperatures within a microscopic theory. Focusing mostly on weakly interacting bosons, we derive an analytical result for the friction force and uncover new regimes of the impurity dynamics. Particularly interesting is the low-temperature $T^2$ dependence of the friction force obtained for a strongly coupled impurity, which should be contrasted with the expected $T^4$ scaling. This new regime applies to systems of bosons with an arbitrary repulsion strength.
		We finally study the evolution of the impurity with a given initial momentum. We evaluate analytically its non-stationary momentum distribution function. The impurity relaxation towards the equilibrium is a realization of the Ornstein-Uhlenbeck process in momentum space. 
	\end{abstract}

	\maketitle
	
	%%%%%%%%%%%%%%%%%%%%%%	
	\emph{Introduction.---} Advances in ultracold quantum gases have led to experiments with unprecedented control over the creation, manipulation and high-resolution imaging of distinguishable particles (impurities) \cite{PhysRevLett.85.483,PhysRevLett.103.150601,PhysRevLett.102.230402,PhysRevA.85.023623,PhysRevLett.109.235301,fukuhara2013quantum,PhysRevLett.117.055302,PhysRevLett.117.055301,Cetina96,Meinert945}. 
	They have opened a new door into a rich physics of the dynamics of an impurity through a many-body environment. A particularly interesting aspect is the possibility to tune the strength of the impurity-environment coupling and explore strong-coupling regimes. 
	
	A one-dimensional quantum liquid is especially interesting environment due to constraints imposed by the reduced dimensionality. The motion of mobile impurities through such a system has attracted a lot of attention \cite{gangardt2009bloch,burovski2014momentum,knap2014quantum,PhysRevA.92.023623,robinson2016motion,mistakidis_dissipative_2019,PhysRevB.101.104503}. A slow impurity propagates without dissipation at zero temperature  \cite{landau+49,PhysRevB.53.9713,PhysRevLett.108.207001}. This hallmark of superfluidity disappears at nonzero temperatures. As a result of impurity scattering off thermally excited quasiparticles, the dissipation occurs. 
	In an early study \cite{PhysRevB.53.9713}, a heavy impurity moving through a Luttinger liquid was considered. It was shown that the friction force experienced by the impurity behaves as the fourth power of temperature. Later, the prefactor of this law has been expressed in terms of the chemical potential of the liquid, the chemical potential of the impurity and its effective mass \cite{gangardt2009bloch,matveev2012scattering}, which are, in general, unknown quantities. This phenomenological theory is valid for arbitrary interaction strengths but is limited to very low temperatures.   
	
	\begin{figure}
		\includegraphics[width=\columnwidth]{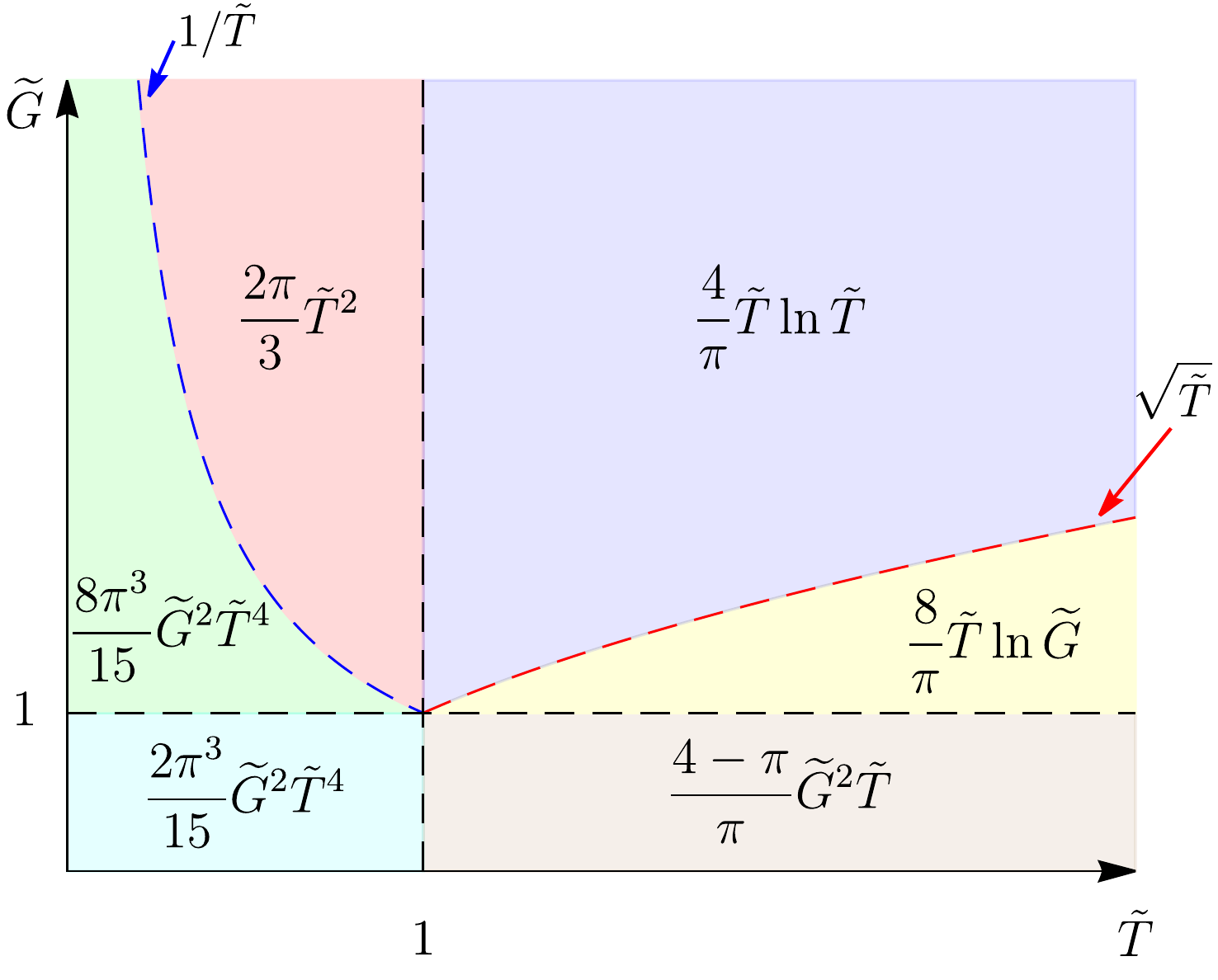}
		\caption{Schematic diagram showing different regimes of the impurity dynamics characterized by the  dimensionless friction  force $\mathcal{F}(\tilde{T},\wt G)$, see Eq.~(\ref{dimensionlessForce}).  The dimensionless temperature is $\tilde{T}=T/m v^2$ and the dimensionless impurity strength is $\wt G=G/\hbar v$. Here $v$ is the sound velocity and $m$ denotes the mass of bosons. The two regions with $T^4$ scaling follow from Eq.~(\ref{GT5}).}
		\label{Fig:contour}
	\end{figure}
	
	In this paper, we significantly advance the state of the art by developing a microscopic theory for the motion of a heavy impurity through a Bose gas in a wide range of temperatures. Our study applies to an arbitrary strong impurity-boson coupling and weak boson-boson interaction.
	We evaluate analytically the friction force and the impurity diffusion constant in momentum space. This enables us to study the relaxation of the impurity with a given initial momentum and find its exact non-stationary distribution function. 
	The friction force shows rich behavior, see Fig.~\ref{Fig:contour}. A particularly striking result is the severe suppression of the parameter region where the $T^4$ dependence occurs and the appearance of $T^2$ law at strong impurity-boson coupling. We argue that this new low-temperature regime exists at arbitrary boson-boson interaction strength. 
	
	Simple understanding of different regimes is as follows. The contribution to the friction force originating from a scattering of a quasiparticle off the impurity is proportional to its reflection probability. The quasiparticles are almost completely reflected off a strongly coupled impurity, while the reflection probability is proportional to the square of the quasiparticle momentum otherwise \cite{KovrzinMaximov}. 
	This behavior provides the explanation for the crossover from $T^4$ into $T^2$ dependence with the increasing impurity coupling: since the momentum of thermally excited quasiparticles is $p \lesssim T/v$,  the exponent of the temperature diminishes by two. Here $v$ denotes the sound velocity. This scenario is realized for an arbitrary strength of the repulsion between the bosons. Moreover, the $T^2$ behavior of the friction force is the impurity-coupling independent, as the refection amplitude is.  Note that the latter regime is out of reach of the theory developed in  Refs.~\cite{gangardt2009bloch,matveev2012scattering} that is a priori limited  to $T\to 0$. 
	
	%%%%%%%%%%%%%%%%%%%%%
	\emph{Model.---}
	%%%%%%%%%%%%%%%%%%%%%
	We study the system of one-dimensional interacting bosons in the presence of a mobile impurity. The bosons are modeled by the Lieb-Liniger Hamiltonian
	\begin{align}\label{Hbosons}
		H_\mathrm{b}=\int \dd x\left(-\hat\psi^\da \dfrac{\hbar^2\partial_x^2}{2m}\hat\psi+\dfrac{g}{2}\hat\psi^\da\hat\psi^\da\hat\psi\hat\psi\right).
	\end{align}
	Here $g>0$ denotes  the repulsive contact interaction between bosons and $m$ is their mass. The bosonic field operators satisfy the commutation relations $[\hat\psi(x),\hat\psi^\da(x')]=\delta(x-x')$ and $[\hat\psi(x),\hat\psi(x')]=0$, with $\delta(x)$ being the Dirac delta function. 
	The total Hamiltonian is given by 
	\begin{align}\label{Hamiltonian}
		H=H_{\mathrm{b}}-\int\dd x\hat\Psi^\da\frac{\hbar^2\partial_x^2}{2M} \hat\Psi+G\int \dd x \hat\Psi^\da \hat\Psi \hat\psi^\da\hat\psi.
	\end{align}
	The second term of Eq.~(\ref{Hamiltonian}) is the kinetic energy of the impurity of the mass $M$ expressed in terms of the impurity field operator $\hat\Psi(x)$. The impurity is coupled to the bosons via the density-density interaction of the strength $G$. We consider a heavy impurity, with a finite mass $M\gg m$. Note that this is very different from the case of an infinitely heavy impurity considered in Refs.~\cite{KaneAndFisher,PhysRevE.55.2835,PhysRevA.66.013610,astrakharchik2004motion,Cherny2012,sykes}. 
	We study weakly interacting bosons that are characterized by the dimensionless parameter  $\gamma=gm/\hbar^2n_0\ll 1$. In this case the quasiparticles of the Hamiltonian (\ref{Hbosons}) have the Bogoliubov dispersion relation $\vep_p=\sqrt{v^2 p^2+p^4/4 m^2}$. Here the sound velocity is $v=\sqrt{gn_0/m}$ and the mean density of the bosons is $n_0$. We introduce the dimensionless parameter $\wt G=G/\hbar v$ that measures the strength of the impurity-boson coupling constant. We consider $\wt{G}$ of an arbitrary strength in the case of repulsion ($\wt{G}>0$), while for an attractive impurity-boson interaction, we restrict our analysis to moderate values $|\wt{G}|\lesssim 1$ due to a collapse of the bosons onto the impurity at strong attraction.  
	
	\emph{Friction force.---} We study the impurity with a velocity smaller than the critical velocity of the superfluid \cite{PhysRevLett.108.207001}, such that it can not emit the Bogoliubov quasiparticles. However, the impurity dissipates energy due to scattering off thermally excited quasiparticles \cite{PhysRevB.53.9713}. The resulting friction force exerted on the impurity can be expressed as the mean change of its momentum in unit time 
	\begin{align}\label{Friction}
		F=\sum_{\delta Q}\delta Q\; W_{Q,Q+\delta Q}.
	\end{align}
	Here $W_{Q,Q+\delta Q}$ denotes the probability per unit time for the impurity to experience a scattering event that changes its momentum from $Q$ into $Q+\delta Q$. 
	
	In order to evaluate $W_{Q,Q+\delta Q}$, let us consider the frame of reference where the impurity is at rest and where an incoming quasiparticle with momentum $p$ scatters off it. We can express $p=\frac{\delta Q}{2}\left[ 1+\mathcal{O}\left(\frac{m}{M}\right)\right]$ using the energy and the momentum conservation laws. Assuming that there is a single quasiparticle with momentum $p$ in the system, its probability to scatter off the impurity per unit time is $|r(p/mv,\wt G)|^2|{\dd \vep_p}/{\dd p}|/L$. Here, $r(p/mv,\wt G)$ denotes the reflection amplitude, the quasiparticle velocity is ${\dd \vep_p}/{\dd p}$, and $L$ is the length of the system. Taking into account that the mean number of quasiparticles of a given energy is determined by the Bose occupation number $n(\epsilon)=(e^{\epsilon/T}-1)^{-1}$,
	we express \cite{PhysRevA.60.3220,PhysRevA.95.053604}
	\begin{align}\label{W}
		W_{Q,Q+\delta Q}=\frac{1}{L}\left|\frac{\dd \vep_p}{\dd p}\right||r(p/mv,\wt G)|^2  n(\tilde{\vep}_p) [1+n(\tilde{\vep}_{p-\delta Q})].
	\end{align} 
	We have assumed that the bosons are in the thermal equilibrium. The energy of the bosonic quasiparticle that scatters off the heavy impurity is $\tilde{\vep}_p=\vep_p+\frac{Q}{M}p$ in the laboratory frame. The Boltzmann constant is set to unity.
	
	Now we are ready to evaluate the friction force by substituting Eq.~(\ref{W}) into Eq.~(\ref{Friction}). The leading order term of the product of the Bose occupation factors appearing in Eq.~(\ref{W}) scales as $(m/M)^0$ and is an even function of $\delta Q$. Since the remaining part of the summand in Eq.~(\ref{Friction}) is an odd function of $\delta Q$, the leading contribution in Eq.~(\ref{Friction}) originates from the term linear in $m/M$ of the product of occupation factors while the rest can be evaluated at $(m/M)^0$ order. We thus obtain
	\begin{align}\label{FrictionF}
		F&=-\frac{m^2 v^2 Q}{\hbar M} \mathcal{F}\left(T/mv^2,\wt G\right),
		\end{align}
	where the dimensionless force is given by
	\begin{align} \label{dimensionlessForce}
		\mathcal{F}(\tilde{T},\wt G)&=\frac{1}{2 \pi \tilde{T}}\int_0^{\infty} \dd k \frac{k^2 \;|r({k},\wt G)|^2}{\sinh^2{\left(k\frac{\sqrt{4+k^2}}{4 \tilde{T}}\right)}}  \frac{2+k^2}{\sqrt{4+k^2}}.
	\end{align}
	We have checked that only the leading term in the expression for $p$ can be used in Eq.~(\ref{Friction}), i.e., $p=\delta Q/2$.
	The reflection amplitude of the Bogoliubov quasiparticles scattering off a static infinitely heavy impurity was  studied in  Refs.~\cite{KovrzinMaximov,Kovrzin,PhysRevA.105.043305} within the Bogoliubov-de Gennes theory. It reads as \cite{PhysRevA.105.043305}
	\begin{widetext}
		\begin{align}
			r(k,\wt G)=\frac{i k \left(1-\eta ^2\right) \left[k^2 (2 \eta +q)+4 \left(\eta ^3+\eta \right)+q^3+2 \eta  q^2+4 \eta ^2 q\right]}{\left[k \eta +i \left(\eta ^2+1\right)\right] (k-i q) \left[k^2 (2 \eta +q)+i k (2 \eta +q)^2-2 \eta  \left(2 \eta ^2+q^2+2 \eta  q-2\right)\right]}\label{r},
		\end{align}
	\end{widetext}
	where $q=\sqrt{4+k^2}$ and $\eta=\left(-\wt G+\sqrt{4+\wt G^2}\right)/2$. 
	In the following, we evaluate analytically the friction force (\ref{FrictionF}) supplemented by Eq.~(\ref{r}) in different parameter regions.
	
	\emph{Strong-coupling regime and low temperatures.---}  At low temperatures, $\tilde{T}=T/mv^2\ll 1$, the low-energy quasiparticles with  momenta $p\lesssim  m v \tilde{T}$ are excited. Then the dimensionless friction force (\ref{dimensionlessForce}) becomes 
	\begin{align}\label{lowT}
		\mathcal{F}(\tilde{T},\wt{G})=\frac{4\tilde{T}^2}{\pi}\int_{0}^{\infty}\frac{x^2 |r(2\tilde{T} x,\wt{G})|^2}{\sinh^2{(x)}} dx.
	\end{align}
	For strong  impurity-boson interaction, $\wt G\gg1$, Eq.~(\ref{r}) becomes
	\begin{align}\label{reflectionSmallP}
		r(p/mv,\wt{G})=\frac{\wt G p}{\wt G p+i mv}.
	\end{align}  
	We distinguish two parameter regions. The reflection probability of quasiparticles with a very low momentum $p\ll mv/ \wt{G}$ is small, $|r|^2=(\wt{G} p/mv)^2$. From Eq.~(\ref{reflectionSmallP}), it follows that the quasiparticles with momentum $p\gg mv/\wt{G}$ are completely reflected off the strongly coupled impurity, $r=1$. Therefore, the friction force shows two different behaviors. For $1/\wt{G}\ll\tilde{T}\ll1$, it becomes $\wt{G}$-independent in the leading order and reads as 
	\begin{align}\label{GT1}
		F=-\frac{2\pi}{3}\frac{T^2Q}{\hbar Mv^2},
	\end{align}
	while at lowest temperatures $\tilde{T}\ll1/\wt{G}\ll1$, it is
	\begin{align}\label{GT2}
		F=-\frac{8\pi^3}{15}\frac{G^2T^4Q}{\hbar^3m^2Mv^8}.
	\end{align}
Thus, we have determined the full dependence of the expected $T^4$ scaling and its parameter region of validity.	
	The crossover between the regimes (\ref{GT1}) and (\ref{GT2}) is given by Eqs.~(\ref{lowT}) and (\ref{reflectionSmallP}). At extremely low temperatures $\tilde{T}\ll \gamma^{1/4}$, the quasiparticles with very small momenta $p\lesssim \gamma^{1/4} mv$ are excited. They are different from the Bogoliubov quasiparticles \cite{PhysRevB.89.100504} and the evaluation of their reflection amplitude requires a separate study \cite{gangardt2009bloch,matveev2012scattering,Schecter_2016}.
	
	It is interesting to note that the new law (\ref{GT1}) should apply for any repulsion strength between the bosons. The reason is that at low temperatures, Eq.~(\ref{lowT}) is valid for any $\gamma$, since the lowest energy excitations are phonons with the dispersion relation $\varepsilon_p=v |p|$. The evaluation of their reflection amplitude $r$ is beyond the scope of this work. However, for a sufficiently strong coupling $\wt{G}$ at a given temperature $\tilde{T}$, the reflection amplitude should approach unity and therefore Eq.~(\ref{GT1}) arises.

	\emph{Strong-coupling regime and high temperatures.---}
	Here, we study in more details the parameter region $\wt G\gg1$ and $\tilde{T}\gg 1$. Then, the quasiparticles with momenta $p\lesssim mv\sqrt{\tilde{T}}$ are thermally excited. 
	We first analyze the behavior of their reflection probability as a function of momentum. From Eq.~(\ref{reflectionSmallP}), it follows that $|r|^2$ increases from zero and reaches one at $p\gg mv/\wt{G}$. It remains one and only at very high momenta, $p\gg mv \wt{G}$, it decays to zero. Indeed, the reflection amplitude (\ref{r}) takes the form 
	$
	r(p/mv,\wt{G})={\wt{G} m v}/{(\wt{G} mv-i p)}
	$  
	for $p\gg mv$. As a result, the dimensionless friction force can be written as 
	\begin{align}\label{crossover2}
		\mathcal{F}(\tilde{T},\wt G)=\frac{2}{\pi}\wt{G}^2\int_0^{\infty}\frac{\dd x}{\sinh^2{(x)}}\frac{\sqrt{1+4x^2\tilde{T}^2}-1}{2\sqrt{1+4x^2\tilde{T}^2}+\wt{G}^2}.
	\end{align}
	We can evaluate the integral (\ref{crossover2}) in the parameter region $\sqrt{\tilde{T}}\gg\wt{G}\gg1$, finding
	\begin{align}\label{GT3}
		F=-\frac{8}{\pi}\frac{mQT}{\hbar M}\ln{(G/\hbar v)}.
	\end{align}
	For $\wt{G}\gg\sqrt{\tilde{T}}\gg1$, Eq.~(\ref{crossover2}) yields the impurity-coupling independent result
	\begin{align}\label{GT4}
		F=-\frac{4}{\pi}\frac{mQT}{\hbar M} \ln{(T/mv^2)}.
	\end{align}
	Note that for an arbitrary $\tilde{T}$, in the case where the impurity coupling constant satisfies $\wt{G}\gg\textrm{max}\{1,\tilde{T}^{-1},\sqrt{\tilde{T}}\}$, after the replacement
	$2\sqrt{1+4x^2\tilde{T}^2}+\wt{G}^2\approx  \wt{G}^2$ in Eq.~(\ref{crossover2}), this modified integral provides the crossover between the high-temperature result (\ref{GT4}) and the low-temperature result (\ref{GT1}). Throughout the paper, we assume $\tilde{T}\ll 1/\sqrt{\gamma}$ such that the system of bosons is in a quantum coherent regime \cite{PhysRevLett.91.040403}. Also the temperature must be sufficiently low such that the impurity is not thermally excited above the critical velocity. For a heavy impurity, the critical velocity can be estimated  to be of the order of $\pi\hbar n_0/M$ \cite{Schecter_2016} leading to the condition  $\tilde{T}\ll\mathrm{min}\{m/M \gamma,1/\sqrt{\gamma}\}$.

	\emph{Other regimes.---} Let us evaluate the friction force in the remaining regimes shown in Fig.~\ref{Fig:contour}.
	At lowest momenta,  $p/mv\ll \textrm{min}\{1,1/\wt{G}\}$, the reflection amplitude (\ref{r}) behaves as  $r(p/mv,\wt{G})\sim p$ . Thus, at $\tilde{T}\ll 1$ and $\wt{G}\tilde{T}\ll1$, we can perform the integration in Eq.~(\ref{lowT}) and obtain 
	 \begin{align}\label{GT5}
	 	F=-\frac{2\pi^3}{15}\frac{QT^4}{\hbar m^2Mv^6}\left(\frac{2+\wt{G}^2}{\sqrt{4+\wt{G}^2}}+\wt{G}-1\right)^2.
	 \end{align}
Here, the impurity potential can be also attractive, satisfying $|\wt{G}|\lesssim 1$. In the case of weak coupling, $|\wt{G}|\ll1$, Eq.~(\ref{GT5}) simplifies into
	$
		F=-{2\pi^3}{G^2T^4Q}/{15\hbar^3m^2Mv^8},
	$
	in agreement with Refs.~\cite{matveev2012scattering,Schecter_2016,PRLimpurity}. The later equation remains valid at extremely low temperatures   $\tilde{T}\ll\gamma^{1/4}$, contrary to Eq.~(\ref{GT2}). Also note that our theory is an expansion in small $m/M$, and thus it  is limited to $G/g\ll M/m$.  Finally, as must be the case, in the strong coupling regime $\tilde{T}^{-1}\gg\wt{G}\gg1$, Eq.~(\ref{GT5}) simplifies to Eq.~(\ref{GT2}).
	
	At high temperatures $\tilde{T}\gg 1$ and $\wt{G}\ll \sqrt{\tilde{T}}$, the dimensionless force (\ref{dimensionlessForce}) becomes linear in $\tilde{T}$
	\begin{align}\label{GT6}
		\mathcal{F}(\tilde{T},\wt{G})=\frac{8\tilde{T}}{\pi} \int_0^{\infty}\dd k\frac{2+k^2}{(4+k^2)^{3/2}}|r({k},\wt G)|^2.
	\end{align}
	 This expression is well defined, since $|r(k,\wt G)|^2\sim 1/k^2$ at large $k$. Note that Eq.~(\ref{GT6}) applies also to the case of attractive impurity potential with $|\wt{G}|\lesssim 1$. At weak coupling $|\wt{G}|\ll 1$, the reflection amplitude simplifies and the integral (\ref{GT6}) can be evaluated analytically. We find 
	$
		F=-{(4-\pi)}{mG^2TQ}/{\pi\hbar^3 Mv^2},
	$
	in agreement with Ref.~\cite{PRLimpurity}. 
	The crossover from the low-temperature  to the high-temperature expression at weak coupling is given by Eq.~(\ref{dimensionlessForce}), once the reflection amplitude is replaced by $|r(p/mv,\wt{G})|= mv{|\wt{G}\;p|  }/{(2m^2v^2+p^2)}.
	$

	\emph{Diffusion coefficient in momentum space.---} 
	Next we evaluate the impurity diffusion coefficient in momentum space, $D=\sum_{\delta Q}\delta Q^2 W_{Q,Q+\delta Q}$. 
	In the leading order, the product of the Bose occupation factors appearing in $W_{Q,Q+\delta Q}$, see Eq.~(\ref{W}), behaves as $(m/M)^0$ and is an even function of $\delta Q$. Since the remaining part of $\delta Q^2 W_{Q,Q+\delta Q}$ is also even, the diffusion coefficient in the leading order satisfies
	\begin{align}\label{D-F}
		D=-2FMT/Q,
	\end{align}
	where $F$ is given by Eq.~(\ref{FrictionF}). Thus, all the above obtained results for the friction force in different parameter regions lead to expressions for the diffusion coefficient, once the relation (\ref{D-F}) is used.
	
	\emph{Impurity relaxation.---} The impurity motion through the quantum liquid is stochastic due to its collisions with thermally excited bosons. In order to characterize it, we consider time dependent distribution function $f(t,Q)$ of the impurity. It satisfies the transport equation 
	\begin{align}\label{kineticE}
		\frac{\partial f(t,Q)}{\partial t}=\sum_{Q'}[f(t,Q')W_{Q',Q}-f(t,Q) W_{Q,Q'}].
	\end{align}
	The scattering probability per unit time $W_{Q,Q+\delta{Q}}$ is a peaked function of $\delta Q$. Since the impurity is heavy, $W_{Q,Q+\delta{Q}}$ varies slowly with $Q$ on the scale of the typical value of $ \delta Q$. Assuming that the distribution function satisfies the latter condition as well, we can bring the collision integral to the Fokker-Planck form \cite{Lifschitz_Pitaevskii_X,Kampen}
	\begin{align}\label{Fokker-Planck}
		\frac{\partial f}{\partial t}=\frac{\partial}{\partial Q}\left[-F f+\frac{1}{2}\frac{\partial(D f)}{\partial Q} \right].
	\end{align}
	The relation (\ref{D-F}) guarantees that the right hand side of Eq.~(\ref{Fokker-Planck}) vanishes once the stationary Maxwell-Boltzmann distribution is reached.  
	
	We can now study the evolution of the distribution function towards the equilibrium one. The initial momentum of the impurity is $Q_0$, i.e.~, $f(0,Q)=\delta(Q-Q_0)$.  In the leading order, the diffusion coefficient characterizing the impurity is $Q$-independent and the friction force is linear in $Q$. Thus, Eq.~(\ref{Fokker-Planck}) describes the well-know Ornstein-Uhlenbeck process  and it can be solved exactly using the method of characteristics. The non-stationary solution reads as \cite{Risken}
	\begin{equation}\label{eq:distribution}
		f(t,Q)=\frac{1}{\sqrt{2\pi MT (1-e^{-2t/\tau})}} \exp{\left[-\frac{(Q-Q_0 e^{-t/\tau})^2}{2MT(1-e^{-2t/\tau})}\right]},
	\end{equation} 
where the relaxation time $\tau=Q/|F|$ is $Q$-independent. $F$ denotes the friction force (\ref{FrictionF}) that we have evaluated previously, see Fig.~\ref{Fig:contour}. 
Equation (\ref{eq:distribution}) enables us to evaluate various statistical properties of the impurity. For example, the average momentum vanishes as
	$
	\overline{Q}=Q_0 \exp{(-t/\tau)}$, while the variance increases as
	$\overline{(Q-\overline{Q})^2}=M T[1- \exp{(-2t/\tau)}]$.
	At short times $t\ll\tau$, the impurity motion in momentum space is diffusive: $\overline{(Q-\overline{Q})^2}= D t$. In the opposite case $t\gg \tau$, the stationary Maxwell-Boltzmann distribution is reached.

	\emph{Conclusions.---} 	
	In this paper, the dissipative dynamics of an impurity in a one-dimensional Bose gas is studied analytically. We have developed a microscopic theory that treats an arbitrarily strong impurity-boson coupling as well as a wide range of temperatures. Various regimes of the impurity motion are shown in Fig.~\ref{Fig:contour}. The striking result is the low-temperature $T^2$ dependence (\ref{GT1}) of the friction force exerted on the impurity. At strong impurity coupling, the region with expected $T^4$ scaling \cite{PhysRevB.53.9713} is strongly suppressed. The reason is that in the presence of a strong impurity coupling, the thermally excited quasiparticles ($p\lesssim \tilde{T}mv$) with momenta above $mv/\widetilde G$ are almost completely reflected from the impurity. Thus, at $\wt{G}\gg 1/\tilde{T}$, the new regime appears. We have argued that it exists for an arbitrary boson-boson repulsion. 
The friction force controls the impurity dynamics. It enters the Fokker-Planck equation as the main parameter. We find its exact solution (\ref{eq:distribution}) describing the relaxation of the impurity with a given initial momentum towards the equilibrium. The experimental setups with cold atoms provide promising test-beds for our predictions due to tunability of the interaction strength between the impurity and host atom through Feshbach resonances and developed powerful measurement techniques \cite{PhysRevLett.85.483,PhysRevLett.103.150601,chin2010feshbach,PhysRevLett.109.235301,PhysRevA.85.023623,Meinert945}. 	
	
	%\bibliography{bib}

\begin{thebibliography}{40}%
		\makeatletter
		\providecommand \@ifxundefined [1]{%
			\@ifx{#1\undefined}
		}%
		\providecommand \@ifnum [1]{%
			\ifnum #1\expandafter \@firstoftwo
			\else \expandafter \@secondoftwo
			\fi
		}%
		\providecommand \@ifx [1]{%
			\ifx #1\expandafter \@firstoftwo
			\else \expandafter \@secondoftwo
			\fi
		}%
		\providecommand \natexlab [1]{#1}%
		\providecommand \enquote  [1]{``#1''}%
		\providecommand \bibnamefont  [1]{#1}%
		\providecommand \bibfnamefont [1]{#1}%
		\providecommand \citenamefont [1]{#1}%
		\providecommand \href@noop [0]{\@secondoftwo}%
		\providecommand \href [0]{\begingroup \@sanitize@url \@href}%
		\providecommand \@href[1]{\@@startlink{#1}\@@href}%
		\providecommand \@@href[1]{\endgroup#1\@@endlink}%
		\providecommand \@sanitize@url [0]{\catcode `\\12\catcode `\$12\catcode
			`\&12\catcode `\#12\catcode `\^12\catcode `\_12\catcode `\%12\relax}%
		\providecommand \@@startlink[1]{}%
		\providecommand \@@endlink[0]{}%
		\providecommand \url  [0]{\begingroup\@sanitize@url \@url }%
		\providecommand \@url [1]{\endgroup\@href {#1}{\urlprefix }}%
		\providecommand \urlprefix  [0]{URL }%
		\providecommand \Eprint [0]{\href }%
		\providecommand \doibase [0]{http://dx.doi.org/}%
		\providecommand \selectlanguage [0]{\@gobble}%
		\providecommand \bibinfo  [0]{\@secondoftwo}%
		\providecommand \bibfield  [0]{\@secondoftwo}%
		\providecommand \translation [1]{[#1]}%
		\providecommand \BibitemOpen [0]{}%
		\providecommand \bibitemStop [0]{}%
		\providecommand \bibitemNoStop [0]{.\EOS\space}%
		\providecommand \EOS [0]{\spacefactor3000\relax}%
		\providecommand \BibitemShut  [1]{\csname bibitem#1\endcsname}%
		\let\auto@bib@innerbib\@empty
		%</preamble>
		\bibitem [{\citenamefont {Chikkatur}\ \emph {et~al.}(2000)\citenamefont
			{Chikkatur}, \citenamefont {G\"orlitz}, \citenamefont {Stamper-Kurn},
			\citenamefont {Inouye}, \citenamefont {Gupta},\ and\ \citenamefont
			{Ketterle}}]{PhysRevLett.85.483}%
		\BibitemOpen
		\bibfield  {author} {\bibinfo {author} {\bibfnamefont {A.~P.}\ \bibnamefont
				{Chikkatur}}, \bibinfo {author} {\bibfnamefont {A.}~\bibnamefont
				{G\"orlitz}}, \bibinfo {author} {\bibfnamefont {D.~M.}\ \bibnamefont
				{Stamper-Kurn}}, \bibinfo {author} {\bibfnamefont {S.}~\bibnamefont
				{Inouye}}, \bibinfo {author} {\bibfnamefont {S.}~\bibnamefont {Gupta}}, \
			and\ \bibinfo {author} {\bibfnamefont {W.}~\bibnamefont {Ketterle}},\ }\href
		{\doibase 10.1103/PhysRevLett.85.483} {\bibfield  {journal} {\bibinfo
				{journal} {Phys. Rev. Lett.}\ }\textbf {\bibinfo {volume} {85}},\ \bibinfo
			{pages} {483} (\bibinfo {year} {2000})}\BibitemShut {NoStop}%
		\bibitem [{\citenamefont {Palzer}\ \emph {et~al.}(2009)\citenamefont {Palzer},
			\citenamefont {Zipkes}, \citenamefont {Sias},\ and\ \citenamefont
			{K\"ohl}}]{PhysRevLett.103.150601}%
		\BibitemOpen
		\bibfield  {author} {\bibinfo {author} {\bibfnamefont {S.}~\bibnamefont
				{Palzer}}, \bibinfo {author} {\bibfnamefont {C.}~\bibnamefont {Zipkes}},
			\bibinfo {author} {\bibfnamefont {C.}~\bibnamefont {Sias}}, \ and\ \bibinfo
			{author} {\bibfnamefont {M.}~\bibnamefont {K\"ohl}},\ }\href {\doibase
			10.1103/PhysRevLett.103.150601} {\bibfield  {journal} {\bibinfo  {journal}
				{Phys. Rev. Lett.}\ }\textbf {\bibinfo {volume} {103}},\ \bibinfo {pages}
			{150601} (\bibinfo {year} {2009})}\BibitemShut {NoStop}%
		\bibitem [{\citenamefont {Schirotzek}\ \emph {et~al.}(2009)\citenamefont
			{Schirotzek}, \citenamefont {Wu}, \citenamefont {Sommer},\ and\ \citenamefont
			{Zwierlein}}]{PhysRevLett.102.230402}%
		\BibitemOpen
		\bibfield  {author} {\bibinfo {author} {\bibfnamefont {A.}~\bibnamefont
				{Schirotzek}}, \bibinfo {author} {\bibfnamefont {C.-H.}\ \bibnamefont {Wu}},
			\bibinfo {author} {\bibfnamefont {A.}~\bibnamefont {Sommer}}, \ and\ \bibinfo
			{author} {\bibfnamefont {M.~W.}\ \bibnamefont {Zwierlein}},\ }\href {\doibase
			10.1103/PhysRevLett.102.230402} {\bibfield  {journal} {\bibinfo  {journal}
				{Phys. Rev. Lett.}\ }\textbf {\bibinfo {volume} {102}},\ \bibinfo {pages}
			{230402} (\bibinfo {year} {2009})}\BibitemShut {NoStop}%
		\bibitem [{\citenamefont {Catani}\ \emph {et~al.}(2012)\citenamefont {Catani},
			\citenamefont {Lamporesi}, \citenamefont {Naik}, \citenamefont {Gring},
			\citenamefont {Inguscio}, \citenamefont {Minardi}, \citenamefont {Kantian},\
			and\ \citenamefont {Giamarchi}}]{PhysRevA.85.023623}%
		\BibitemOpen
		\bibfield  {author} {\bibinfo {author} {\bibfnamefont {J.}~\bibnamefont
				{Catani}}, \bibinfo {author} {\bibfnamefont {G.}~\bibnamefont {Lamporesi}},
			\bibinfo {author} {\bibfnamefont {D.}~\bibnamefont {Naik}}, \bibinfo {author}
			{\bibfnamefont {M.}~\bibnamefont {Gring}}, \bibinfo {author} {\bibfnamefont
				{M.}~\bibnamefont {Inguscio}}, \bibinfo {author} {\bibfnamefont
				{F.}~\bibnamefont {Minardi}}, \bibinfo {author} {\bibfnamefont
				{A.}~\bibnamefont {Kantian}}, \ and\ \bibinfo {author} {\bibfnamefont
				{T.}~\bibnamefont {Giamarchi}},\ }\href {\doibase 10.1103/PhysRevA.85.023623}
		{\bibfield  {journal} {\bibinfo  {journal} {Phys. Rev. A}\ }\textbf {\bibinfo
				{volume} {85}},\ \bibinfo {pages} {023623} (\bibinfo {year}
			{2012})}\BibitemShut {NoStop}%
		\bibitem [{\citenamefont {Spethmann}\ \emph {et~al.}(2012)\citenamefont
			{Spethmann}, \citenamefont {Kindermann}, \citenamefont {John}, \citenamefont
			{Weber}, \citenamefont {Meschede},\ and\ \citenamefont
			{Widera}}]{PhysRevLett.109.235301}%
		\BibitemOpen
		\bibfield  {author} {\bibinfo {author} {\bibfnamefont {N.}~\bibnamefont
				{Spethmann}}, \bibinfo {author} {\bibfnamefont {F.}~\bibnamefont
				{Kindermann}}, \bibinfo {author} {\bibfnamefont {S.}~\bibnamefont {John}},
			\bibinfo {author} {\bibfnamefont {C.}~\bibnamefont {Weber}}, \bibinfo
			{author} {\bibfnamefont {D.}~\bibnamefont {Meschede}}, \ and\ \bibinfo
			{author} {\bibfnamefont {A.}~\bibnamefont {Widera}},\ }\href {\doibase
			10.1103/PhysRevLett.109.235301} {\bibfield  {journal} {\bibinfo  {journal}
				{Phys. Rev. Lett.}\ }\textbf {\bibinfo {volume} {109}},\ \bibinfo {pages}
			{235301} (\bibinfo {year} {2012})}\BibitemShut {NoStop}%
		\bibitem [{\citenamefont {Fukuhara}\ \emph {et~al.}(2013)\citenamefont
			{Fukuhara}, \citenamefont {Kantian}, \citenamefont {Endres}, \citenamefont
			{Cheneau}, \citenamefont {Schau{\ss}}, \citenamefont {Hild}, \citenamefont
			{Bellem}, \citenamefont {Schollw{\"o}ck}, \citenamefont {Giamarchi},
			\citenamefont {Gross}, \citenamefont {Bloch},\ and\ \citenamefont
			{Kuhr}}]{fukuhara2013quantum}%
		\BibitemOpen
		\bibfield  {author} {\bibinfo {author} {\bibfnamefont {T.}~\bibnamefont
				{Fukuhara}}, \bibinfo {author} {\bibfnamefont {A.}~\bibnamefont {Kantian}},
			\bibinfo {author} {\bibfnamefont {M.}~\bibnamefont {Endres}}, \bibinfo
			{author} {\bibfnamefont {M.}~\bibnamefont {Cheneau}}, \bibinfo {author}
			{\bibfnamefont {P.}~\bibnamefont {Schau{\ss}}}, \bibinfo {author}
			{\bibfnamefont {S.}~\bibnamefont {Hild}}, \bibinfo {author} {\bibfnamefont
				{D.}~\bibnamefont {Bellem}}, \bibinfo {author} {\bibfnamefont
				{U.}~\bibnamefont {Schollw{\"o}ck}}, \bibinfo {author} {\bibfnamefont
				{T.}~\bibnamefont {Giamarchi}}, \bibinfo {author} {\bibfnamefont
				{C.}~\bibnamefont {Gross}}, \bibinfo {author} {\bibfnamefont
				{I.}~\bibnamefont {Bloch}}, \ and\ \bibinfo {author} {\bibfnamefont
				{S.}~\bibnamefont {Kuhr}},\ }\href {\doibase 10.1038/nphys2561} {\bibfield
			{journal} {\bibinfo  {journal} {Nat. Phys.}\ }\textbf {\bibinfo {volume}
				{9}},\ \bibinfo {pages} {235} (\bibinfo {year} {2013})}\BibitemShut {NoStop}%
		\bibitem [{\citenamefont {J\o{}rgensen}\ \emph {et~al.}(2016)\citenamefont
			{J\o{}rgensen}, \citenamefont {Wacker}, \citenamefont {Skalmstang},
			\citenamefont {Parish}, \citenamefont {Levinsen}, \citenamefont
			{Christensen}, \citenamefont {Bruun},\ and\ \citenamefont
			{Arlt}}]{PhysRevLett.117.055302}%
		\BibitemOpen
		\bibfield  {author} {\bibinfo {author} {\bibfnamefont {N.~B.}\ \bibnamefont
				{J\o{}rgensen}}, \bibinfo {author} {\bibfnamefont {L.}~\bibnamefont
				{Wacker}}, \bibinfo {author} {\bibfnamefont {K.~T.}\ \bibnamefont
				{Skalmstang}}, \bibinfo {author} {\bibfnamefont {M.~M.}\ \bibnamefont
				{Parish}}, \bibinfo {author} {\bibfnamefont {J.}~\bibnamefont {Levinsen}},
			\bibinfo {author} {\bibfnamefont {R.~S.}\ \bibnamefont {Christensen}},
			\bibinfo {author} {\bibfnamefont {G.~M.}\ \bibnamefont {Bruun}}, \ and\
			\bibinfo {author} {\bibfnamefont {J.~J.}\ \bibnamefont {Arlt}},\ }\href
		{\doibase 10.1103/PhysRevLett.117.055302} {\bibfield  {journal} {\bibinfo
				{journal} {Phys. Rev. Lett.}\ }\textbf {\bibinfo {volume} {117}},\ \bibinfo
			{pages} {055302} (\bibinfo {year} {2016})}\BibitemShut {NoStop}%
		\bibitem [{\citenamefont {Hu}\ \emph {et~al.}(2016)\citenamefont {Hu},
			\citenamefont {Van~de Graaff}, \citenamefont {Kedar}, \citenamefont {Corson},
			\citenamefont {Cornell},\ and\ \citenamefont {Jin}}]{PhysRevLett.117.055301}%
		\BibitemOpen
		\bibfield  {author} {\bibinfo {author} {\bibfnamefont {M.-G.}\ \bibnamefont
				{Hu}}, \bibinfo {author} {\bibfnamefont {M.~J.}\ \bibnamefont {Van~de
					Graaff}}, \bibinfo {author} {\bibfnamefont {D.}~\bibnamefont {Kedar}},
			\bibinfo {author} {\bibfnamefont {J.~P.}\ \bibnamefont {Corson}}, \bibinfo
			{author} {\bibfnamefont {E.~A.}\ \bibnamefont {Cornell}}, \ and\ \bibinfo
			{author} {\bibfnamefont {D.~S.}\ \bibnamefont {Jin}},\ }\href {\doibase
			10.1103/PhysRevLett.117.055301} {\bibfield  {journal} {\bibinfo  {journal}
				{Phys. Rev. Lett.}\ }\textbf {\bibinfo {volume} {117}},\ \bibinfo {pages}
			{055301} (\bibinfo {year} {2016})}\BibitemShut {NoStop}%
		\bibitem [{\citenamefont {Cetina}\ \emph {et~al.}(2016)\citenamefont {Cetina},
			\citenamefont {Jag}, \citenamefont {Lous}, \citenamefont {Fritsche},
			\citenamefont {Walraven}, \citenamefont {Grimm}, \citenamefont {Levinsen},
			\citenamefont {Parish}, \citenamefont {Schmidt}, \citenamefont {Knap},\ and\
			\citenamefont {Demler}}]{Cetina96}%
		\BibitemOpen
		\bibfield  {author} {\bibinfo {author} {\bibfnamefont {M.}~\bibnamefont
				{Cetina}}, \bibinfo {author} {\bibfnamefont {M.}~\bibnamefont {Jag}},
			\bibinfo {author} {\bibfnamefont {R.~S.}\ \bibnamefont {Lous}}, \bibinfo
			{author} {\bibfnamefont {I.}~\bibnamefont {Fritsche}}, \bibinfo {author}
			{\bibfnamefont {J.~T.~M.}\ \bibnamefont {Walraven}}, \bibinfo {author}
			{\bibfnamefont {R.}~\bibnamefont {Grimm}}, \bibinfo {author} {\bibfnamefont
				{J.}~\bibnamefont {Levinsen}}, \bibinfo {author} {\bibfnamefont {M.~M.}\
				\bibnamefont {Parish}}, \bibinfo {author} {\bibfnamefont {R.}~\bibnamefont
				{Schmidt}}, \bibinfo {author} {\bibfnamefont {M.}~\bibnamefont {Knap}}, \
			and\ \bibinfo {author} {\bibfnamefont {E.}~\bibnamefont {Demler}},\ }\href
		{\doibase 10.1126/science.aaf5134} {\bibfield  {journal} {\bibinfo  {journal}
				{Science}\ }\textbf {\bibinfo {volume} {354}},\ \bibinfo {pages} {96}
			(\bibinfo {year} {2016})}\BibitemShut {NoStop}%
		\bibitem [{\citenamefont {Meinert}\ \emph {et~al.}(2017)\citenamefont
			{Meinert}, \citenamefont {Knap}, \citenamefont {Kirilov}, \citenamefont
			{Jag-Lauber}, \citenamefont {Zvonarev}, \citenamefont {Demler},\ and\
			\citenamefont {N{\"a}gerl}}]{Meinert945}%
		\BibitemOpen
		\bibfield  {author} {\bibinfo {author} {\bibfnamefont {F.}~\bibnamefont
				{Meinert}}, \bibinfo {author} {\bibfnamefont {M.}~\bibnamefont {Knap}},
			\bibinfo {author} {\bibfnamefont {E.}~\bibnamefont {Kirilov}}, \bibinfo
			{author} {\bibfnamefont {K.}~\bibnamefont {Jag-Lauber}}, \bibinfo {author}
			{\bibfnamefont {M.~B.}\ \bibnamefont {Zvonarev}}, \bibinfo {author}
			{\bibfnamefont {E.}~\bibnamefont {Demler}}, \ and\ \bibinfo {author}
			{\bibfnamefont {H.-C.}\ \bibnamefont {N{\"a}gerl}},\ }\href {\doibase
			10.1126/science.aah6616} {\bibfield  {journal} {\bibinfo  {journal}
				{Science}\ }\textbf {\bibinfo {volume} {356}},\ \bibinfo {pages} {945}
			(\bibinfo {year} {2017})}\BibitemShut {NoStop}%
		\bibitem [{\citenamefont {Gangardt}\ and\ \citenamefont
			{Kamenev}(2009)}]{gangardt2009bloch}%
		\BibitemOpen
		\bibfield  {author} {\bibinfo {author} {\bibfnamefont {D.~M.}\ \bibnamefont
				{Gangardt}}\ and\ \bibinfo {author} {\bibfnamefont {A.}~\bibnamefont
				{Kamenev}},\ }\href {\doibase 10.1103/PhysRevLett.102.070402} {\bibfield
			{journal} {\bibinfo  {journal} {Phys. Rev. Lett.}\ }\textbf {\bibinfo
				{volume} {102}},\ \bibinfo {pages} {070402} (\bibinfo {year}
			{2009})}\BibitemShut {NoStop}%
		\bibitem [{\citenamefont {Burovski}\ \emph {et~al.}(2014)\citenamefont
			{Burovski}, \citenamefont {Cheianov}, \citenamefont {Gamayun},\ and\
			\citenamefont {Lychkovskiy}}]{burovski2014momentum}%
		\BibitemOpen
		\bibfield  {author} {\bibinfo {author} {\bibfnamefont {E.}~\bibnamefont
				{Burovski}}, \bibinfo {author} {\bibfnamefont {V.}~\bibnamefont {Cheianov}},
			\bibinfo {author} {\bibfnamefont {O.}~\bibnamefont {Gamayun}}, \ and\
			\bibinfo {author} {\bibfnamefont {O.}~\bibnamefont {Lychkovskiy}},\ }\href
		{\doibase 10.1103/PhysRevA.89.041601} {\bibfield  {journal} {\bibinfo
				{journal} {Phys. Rev. A}\ }\textbf {\bibinfo {volume} {89}},\ \bibinfo
			{pages} {041601} (\bibinfo {year} {2014})}\BibitemShut {NoStop}%
		\bibitem [{\citenamefont {Knap}\ \emph {et~al.}(2014)\citenamefont {Knap},
			\citenamefont {Mathy}, \citenamefont {Ganahl}, \citenamefont {Zvonarev},\
			and\ \citenamefont {Demler}}]{knap2014quantum}%
		\BibitemOpen
		\bibfield  {author} {\bibinfo {author} {\bibfnamefont {M.}~\bibnamefont
				{Knap}}, \bibinfo {author} {\bibfnamefont {C.~J.~M.}\ \bibnamefont {Mathy}},
			\bibinfo {author} {\bibfnamefont {M.}~\bibnamefont {Ganahl}}, \bibinfo
			{author} {\bibfnamefont {M.~B.}\ \bibnamefont {Zvonarev}}, \ and\ \bibinfo
			{author} {\bibfnamefont {E.}~\bibnamefont {Demler}},\ }\href {\doibase
			10.1103/PhysRevLett.112.015302} {\bibfield  {journal} {\bibinfo  {journal}
				{Phys. Rev. Lett.}\ }\textbf {\bibinfo {volume} {112}},\ \bibinfo {pages}
			{015302} (\bibinfo {year} {2014})}\BibitemShut {NoStop}%
		\bibitem [{\citenamefont {Volosniev}\ \emph {et~al.}(2015)\citenamefont
			{Volosniev}, \citenamefont {Hammer},\ and\ \citenamefont
			{Zinner}}]{PhysRevA.92.023623}%
		\BibitemOpen
		\bibfield  {author} {\bibinfo {author} {\bibfnamefont {A.~G.}\ \bibnamefont
				{Volosniev}}, \bibinfo {author} {\bibfnamefont {H.-W.}\ \bibnamefont
				{Hammer}}, \ and\ \bibinfo {author} {\bibfnamefont {N.~T.}\ \bibnamefont
				{Zinner}},\ }\href {\doibase 10.1103/PhysRevA.92.023623} {\bibfield
			{journal} {\bibinfo  {journal} {Phys. Rev. A}\ }\textbf {\bibinfo {volume}
				{92}},\ \bibinfo {pages} {023623} (\bibinfo {year} {2015})}\BibitemShut
		{NoStop}%
		\bibitem [{\citenamefont {Robinson}\ \emph {et~al.}(2016)\citenamefont
			{Robinson}, \citenamefont {Caux},\ and\ \citenamefont
			{Konik}}]{robinson2016motion}%
		\BibitemOpen
		\bibfield  {author} {\bibinfo {author} {\bibfnamefont {N.~J.}\ \bibnamefont
				{Robinson}}, \bibinfo {author} {\bibfnamefont {J.-S.}\ \bibnamefont {Caux}},
			\ and\ \bibinfo {author} {\bibfnamefont {R.~M.}\ \bibnamefont {Konik}},\
		}\href {\doibase 10.1103/PhysRevLett.116.145302} {\bibfield  {journal}
			{\bibinfo  {journal} {Phys. Rev. Lett.}\ }\textbf {\bibinfo {volume} {116}},\
			\bibinfo {pages} {145302} (\bibinfo {year} {2016})}\BibitemShut {NoStop}%
		\bibitem [{\citenamefont {Mistakidis}\ \emph {et~al.}(2019)\citenamefont
			{Mistakidis}, \citenamefont {Grusdt}, \citenamefont {Koutentakis},\ and\
			\citenamefont {Schmelcher}}]{mistakidis_dissipative_2019}%
		\BibitemOpen
		\bibfield  {author} {\bibinfo {author} {\bibfnamefont {S.~I.}\ \bibnamefont
				{Mistakidis}}, \bibinfo {author} {\bibfnamefont {F.}~\bibnamefont {Grusdt}},
			\bibinfo {author} {\bibfnamefont {G.~M.}\ \bibnamefont {Koutentakis}}, \ and\
			\bibinfo {author} {\bibfnamefont {P.}~\bibnamefont {Schmelcher}},\ }\href
		{\doibase 10.1088/1367-2630/ab4738} {\bibfield  {journal} {\bibinfo
				{journal} {New J. Phys.}\ }\textbf {\bibinfo {volume} {21}},\ \bibinfo
			{pages} {103026} (\bibinfo {year} {2019})}\BibitemShut {NoStop}%
		\bibitem [{\citenamefont {Petkovi\ifmmode~\acute{c}\else
				\'{c}\fi{}}(2020)}]{PhysRevB.101.104503}%
		\BibitemOpen
		\bibfield  {author} {\bibinfo {author} {\bibfnamefont {A.}~\bibnamefont
				{Petkovi\ifmmode~\acute{c}\else \'{c}\fi{}}},\ }\href {\doibase
			10.1103/PhysRevB.101.104503} {\bibfield  {journal} {\bibinfo  {journal}
				{Phys. Rev. B}\ }\textbf {\bibinfo {volume} {101}},\ \bibinfo {pages}
			{104503} (\bibinfo {year} {2020})}\BibitemShut {NoStop}%
		\bibitem [{\citenamefont {Landau}\ and\ \citenamefont
			{Khalatnikov}(1949)}]{landau+49}%
		\BibitemOpen
		\bibfield  {author} {\bibinfo {author} {\bibfnamefont {L.~D.}\ \bibnamefont
				{Landau}}\ and\ \bibinfo {author} {\bibfnamefont {I.~M.}\ \bibnamefont
				{Khalatnikov}},\ }\href@noop {} {\bibfield  {journal} {\bibinfo  {journal}
				{Zh. Eksp. Teor. Fiz.}\ }\textbf {\bibinfo {volume} {19}},\ \bibinfo {pages}
			{709} (\bibinfo {year} {1949})}\BibitemShut {NoStop}%
		\bibitem [{\citenamefont {Castro~Neto}\ and\ \citenamefont
			{Fisher}(1996)}]{PhysRevB.53.9713}%
		\BibitemOpen
		\bibfield  {author} {\bibinfo {author} {\bibfnamefont {A.~H.}\ \bibnamefont
				{Castro~Neto}}\ and\ \bibinfo {author} {\bibfnamefont {M.~P.~A.}\
				\bibnamefont {Fisher}},\ }\href {\doibase 10.1103/PhysRevB.53.9713}
		{\bibfield  {journal} {\bibinfo  {journal} {Phys. Rev. B}\ }\textbf {\bibinfo
				{volume} {53}},\ \bibinfo {pages} {9713} (\bibinfo {year}
			{1996})}\BibitemShut {NoStop}%
		\bibitem [{\citenamefont {Schecter}\ \emph {et~al.}(2012)\citenamefont
			{Schecter}, \citenamefont {Kamenev}, \citenamefont {Gangardt},\ and\
			\citenamefont {Lamacraft}}]{PhysRevLett.108.207001}%
		\BibitemOpen
		\bibfield  {author} {\bibinfo {author} {\bibfnamefont {M.}~\bibnamefont
				{Schecter}}, \bibinfo {author} {\bibfnamefont {A.}~\bibnamefont {Kamenev}},
			\bibinfo {author} {\bibfnamefont {D.~M.}\ \bibnamefont {Gangardt}}, \ and\
			\bibinfo {author} {\bibfnamefont {A.}~\bibnamefont {Lamacraft}},\ }\href
		{\doibase 10.1103/PhysRevLett.108.207001} {\bibfield  {journal} {\bibinfo
				{journal} {Phys. Rev. Lett.}\ }\textbf {\bibinfo {volume} {108}},\ \bibinfo
			{pages} {207001} (\bibinfo {year} {2012})}\BibitemShut {NoStop}%
		\bibitem [{\citenamefont {Matveev}\ and\ \citenamefont
			{Andreev}(2012)}]{matveev2012scattering}%
		\BibitemOpen
		\bibfield  {author} {\bibinfo {author} {\bibfnamefont {K.~A.}\ \bibnamefont
				{Matveev}}\ and\ \bibinfo {author} {\bibfnamefont {A.~V.}\ \bibnamefont
				{Andreev}},\ }\href {\doibase 10.1103/PhysRevB.86.045136} {\bibfield
			{journal} {\bibinfo  {journal} {Phys. Rev. B}\ }\textbf {\bibinfo {volume}
				{86}},\ \bibinfo {pages} {045136} (\bibinfo {year} {2012})}\BibitemShut
		{NoStop}%
		\bibitem [{\citenamefont {Kovrizhin}\ and\ \citenamefont
			{Maksimov}(2001)}]{KovrzinMaximov}%
		\BibitemOpen
		\bibfield  {author} {\bibinfo {author} {\bibfnamefont {D.}~\bibnamefont
				{Kovrizhin}}\ and\ \bibinfo {author} {\bibfnamefont {L.}~\bibnamefont
				{Maksimov}},\ }\href {\doibase https://doi.org/10.1134/1.1378096} {\bibfield
			{journal} {\bibinfo  {journal} {Dokl. Phys.}\ }\textbf {\bibinfo {volume}
				{46}},\ \bibinfo {pages} {328} (\bibinfo {year} {2001})}\BibitemShut
		{NoStop}%
		\bibitem [{\citenamefont {Kane}\ and\ \citenamefont
			{Fisher}(1992)}]{KaneAndFisher}%
		\BibitemOpen
		\bibfield  {author} {\bibinfo {author} {\bibfnamefont {C.~L.}\ \bibnamefont
				{Kane}}\ and\ \bibinfo {author} {\bibfnamefont {M.~P.~A.}\ \bibnamefont
				{Fisher}},\ }\href {\doibase 10.1103/PhysRevB.46.15233} {\bibfield  {journal}
			{\bibinfo  {journal} {Phys. Rev. B}\ }\textbf {\bibinfo {volume} {46}},\
			\bibinfo {pages} {15233} (\bibinfo {year} {1992})}\BibitemShut {NoStop}%
		\bibitem [{\citenamefont {Hakim}(1997)}]{PhysRevE.55.2835}%
		\BibitemOpen
		\bibfield  {author} {\bibinfo {author} {\bibfnamefont {V.}~\bibnamefont
				{Hakim}},\ }\href {\doibase 10.1103/PhysRevE.55.2835} {\bibfield  {journal}
			{\bibinfo  {journal} {Phys. Rev. E}\ }\textbf {\bibinfo {volume} {55}},\
			\bibinfo {pages} {2835} (\bibinfo {year} {1997})}\BibitemShut {NoStop}%
		\bibitem [{\citenamefont {Pavloff}(2002)}]{PhysRevA.66.013610}%
		\BibitemOpen
		\bibfield  {author} {\bibinfo {author} {\bibfnamefont {N.}~\bibnamefont
				{Pavloff}},\ }\href {\doibase 10.1103/PhysRevA.66.013610} {\bibfield
			{journal} {\bibinfo  {journal} {Phys. Rev. A}\ }\textbf {\bibinfo {volume}
				{66}},\ \bibinfo {pages} {013610} (\bibinfo {year} {2002})}\BibitemShut
		{NoStop}%
		\bibitem [{\citenamefont {Astrakharchik}\ and\ \citenamefont
			{Pitaevskii}(2004)}]{astrakharchik2004motion}%
		\BibitemOpen
		\bibfield  {author} {\bibinfo {author} {\bibfnamefont {G.~E.}\ \bibnamefont
				{Astrakharchik}}\ and\ \bibinfo {author} {\bibfnamefont {L.~P.}\ \bibnamefont
				{Pitaevskii}},\ }\href {\doibase 10.1103/PhysRevA.70.013608} {\bibfield
			{journal} {\bibinfo  {journal} {Phys. Rev. A}\ }\textbf {\bibinfo {volume}
				{70}},\ \bibinfo {pages} {013608} (\bibinfo {year} {2004})}\BibitemShut
		{NoStop}%
		\bibitem [{\citenamefont {Cherny}\ \emph {et~al.}(2012)\citenamefont {Cherny},
			\citenamefont {Caux},\ and\ \citenamefont {Brand}}]{Cherny2012}%
		\BibitemOpen
		\bibfield  {author} {\bibinfo {author} {\bibfnamefont {A.~Y.}\ \bibnamefont
				{Cherny}}, \bibinfo {author} {\bibfnamefont {J.-S.}\ \bibnamefont {Caux}}, \
			and\ \bibinfo {author} {\bibfnamefont {J.}~\bibnamefont {Brand}},\ }\href
		{\doibase 10.1007/s11467-011-0211-2} {\bibfield  {journal} {\bibinfo
				{journal} {Front. Phys.}\ }\textbf {\bibinfo {volume} {7}},\ \bibinfo {pages}
			{54} (\bibinfo {year} {2012})}\BibitemShut {NoStop}%
		\bibitem [{\citenamefont {Sykes}\ \emph {et~al.}(2009)\citenamefont {Sykes},
			\citenamefont {Davis},\ and\ \citenamefont {Roberts}}]{sykes}%
		\BibitemOpen
		\bibfield  {author} {\bibinfo {author} {\bibfnamefont {A.~G.}\ \bibnamefont
				{Sykes}}, \bibinfo {author} {\bibfnamefont {M.~J.}\ \bibnamefont {Davis}}, \
			and\ \bibinfo {author} {\bibfnamefont {D.~C.}\ \bibnamefont {Roberts}},\
		}\href {\doibase 10.1103/PhysRevLett.103.085302} {\bibfield  {journal}
			{\bibinfo  {journal} {Phys. Rev. Lett.}\ }\textbf {\bibinfo {volume} {103}},\
			\bibinfo {pages} {085302} (\bibinfo {year} {2009})}\BibitemShut {NoStop}%
		\bibitem [{\citenamefont {Fedichev}\ \emph {et~al.}(1999)\citenamefont
			{Fedichev}, \citenamefont {Muryshev},\ and\ \citenamefont
			{Shlyapnikov}}]{PhysRevA.60.3220}%
		\BibitemOpen
		\bibfield  {author} {\bibinfo {author} {\bibfnamefont {P.~O.}\ \bibnamefont
				{Fedichev}}, \bibinfo {author} {\bibfnamefont {A.~E.}\ \bibnamefont
				{Muryshev}}, \ and\ \bibinfo {author} {\bibfnamefont {G.~V.}\ \bibnamefont
				{Shlyapnikov}},\ }\href {\doibase 10.1103/PhysRevA.60.3220} {\bibfield
			{journal} {\bibinfo  {journal} {Phys. Rev. A}\ }\textbf {\bibinfo {volume}
				{60}},\ \bibinfo {pages} {3220} (\bibinfo {year} {1999})}\BibitemShut
		{NoStop}%
		\bibitem [{\citenamefont {Hurst}\ \emph {et~al.}(2017)\citenamefont {Hurst},
			\citenamefont {Efimkin}, \citenamefont {Spielman},\ and\ \citenamefont
			{Galitski}}]{PhysRevA.95.053604}%
		\BibitemOpen
		\bibfield  {author} {\bibinfo {author} {\bibfnamefont {H.~M.}\ \bibnamefont
				{Hurst}}, \bibinfo {author} {\bibfnamefont {D.~K.}\ \bibnamefont {Efimkin}},
			\bibinfo {author} {\bibfnamefont {I.~B.}\ \bibnamefont {Spielman}}, \ and\
			\bibinfo {author} {\bibfnamefont {V.}~\bibnamefont {Galitski}},\ }\href
		{\doibase 10.1103/PhysRevA.95.053604} {\bibfield  {journal} {\bibinfo
				{journal} {Phys. Rev. A}\ }\textbf {\bibinfo {volume} {95}},\ \bibinfo
			{pages} {053604} (\bibinfo {year} {2017})}\BibitemShut {NoStop}%
		\bibitem [{\citenamefont {Kovrizhin}(2001)}]{Kovrzin}%
		\BibitemOpen
		\bibfield  {author} {\bibinfo {author} {\bibfnamefont {D.}~\bibnamefont
				{Kovrizhin}},\ }\href {\doibase
			https://doi.org/10.1016/S0375-9601(01)00503-5} {\bibfield  {journal}
			{\bibinfo  {journal} {Phys. Lett. A}\ }\textbf {\bibinfo {volume} {287}},\
			\bibinfo {pages} {392} (\bibinfo {year} {2001})}\BibitemShut {NoStop}%
		\bibitem [{\citenamefont {Petkovi\ifmmode~\acute{c}\else
				\'{c}\fi{}}(2022)}]{PhysRevA.105.043305}%
		\BibitemOpen
		\bibfield  {author} {\bibinfo {author} {\bibfnamefont {A.}~\bibnamefont
				{Petkovi\ifmmode~\acute{c}\else \'{c}\fi{}}},\ }\href {\doibase
			10.1103/PhysRevA.105.043305} {\bibfield  {journal} {\bibinfo  {journal}
				{Phys. Rev. A}\ }\textbf {\bibinfo {volume} {105}},\ \bibinfo {pages}
			{043305} (\bibinfo {year} {2022})}\BibitemShut {NoStop}%
		\bibitem [{\citenamefont {Pustilnik}\ and\ \citenamefont
			{Matveev}(2014)}]{PhysRevB.89.100504}%
		\BibitemOpen
		\bibfield  {author} {\bibinfo {author} {\bibfnamefont {M.}~\bibnamefont
				{Pustilnik}}\ and\ \bibinfo {author} {\bibfnamefont {K.~A.}\ \bibnamefont
				{Matveev}},\ }\href {\doibase 10.1103/PhysRevB.89.100504} {\bibfield
			{journal} {\bibinfo  {journal} {Phys. Rev. B}\ }\textbf {\bibinfo {volume}
				{89}},\ \bibinfo {pages} {100504} (\bibinfo {year} {2014})}\BibitemShut
		{NoStop}%
		\bibitem [{\citenamefont {Schecter}\ \emph {et~al.}(2016)\citenamefont
			{Schecter}, \citenamefont {Gangardt},\ and\ \citenamefont
			{Kamenev}}]{Schecter_2016}%
		\BibitemOpen
		\bibfield  {author} {\bibinfo {author} {\bibfnamefont {M.}~\bibnamefont
				{Schecter}}, \bibinfo {author} {\bibfnamefont {D.~M.}\ \bibnamefont
				{Gangardt}}, \ and\ \bibinfo {author} {\bibfnamefont {A.}~\bibnamefont
				{Kamenev}},\ }\href {\doibase 10.1088/1367-2630/18/6/065002} {\bibfield
			{journal} {\bibinfo  {journal} {New J. Phys.}\ }\textbf {\bibinfo {volume}
				{18}},\ \bibinfo {pages} {065002} (\bibinfo {year} {2016})}\BibitemShut
		{NoStop}%
		\bibitem [{\citenamefont {Kheruntsyan}\ \emph {et~al.}(2003)\citenamefont
			{Kheruntsyan}, \citenamefont {Gangardt}, \citenamefont {Drummond},\ and\
			\citenamefont {Shlyapnikov}}]{PhysRevLett.91.040403}%
		\BibitemOpen
		\bibfield  {author} {\bibinfo {author} {\bibfnamefont {K.~V.}\ \bibnamefont
				{Kheruntsyan}}, \bibinfo {author} {\bibfnamefont {D.~M.}\ \bibnamefont
				{Gangardt}}, \bibinfo {author} {\bibfnamefont {P.~D.}\ \bibnamefont
				{Drummond}}, \ and\ \bibinfo {author} {\bibfnamefont {G.~V.}\ \bibnamefont
				{Shlyapnikov}},\ }\href {\doibase 10.1103/PhysRevLett.91.040403} {\bibfield
			{journal} {\bibinfo  {journal} {Phys. Rev. Lett.}\ }\textbf {\bibinfo
				{volume} {91}},\ \bibinfo {pages} {040403} (\bibinfo {year}
			{2003})}\BibitemShut {NoStop}%
		\bibitem [{\citenamefont {Petkovi\ifmmode~\acute{c}\else \'{c}\fi{}}\ and\
			\citenamefont {Ristivojevic}(2016)}]{PRLimpurity}%
		\BibitemOpen
		\bibfield  {author} {\bibinfo {author} {\bibfnamefont {A.}~\bibnamefont
				{Petkovi\ifmmode~\acute{c}\else \'{c}\fi{}}}\ and\ \bibinfo {author}
			{\bibfnamefont {Z.}~\bibnamefont {Ristivojevic}},\ }\href {\doibase
			10.1103/PhysRevLett.117.105301} {\bibfield  {journal} {\bibinfo  {journal}
				{Phys. Rev. Lett.}\ }\textbf {\bibinfo {volume} {117}},\ \bibinfo {pages}
			{105301} (\bibinfo {year} {2016})}\BibitemShut {NoStop}%
		\bibitem [{\citenamefont {Lifshitz}\ and\ \citenamefont
			{Pitaevskii}(1983)}]{Lifschitz_Pitaevskii_X}%
		\BibitemOpen
		\bibfield  {author} {\bibinfo {author} {\bibfnamefont {E.~M.}\ \bibnamefont
				{Lifshitz}}\ and\ \bibinfo {author} {\bibfnamefont {L.~P.}\ \bibnamefont
				{Pitaevskii}},\ }\href@noop {} {\emph {\bibinfo {title} {Physical Kinetics,
					Course of Theoretical Physics}}},\ Vol.~\bibinfo {volume} {10}\ (\bibinfo
		{publisher} {Oxford},\ \bibinfo {year} {1983})\BibitemShut {NoStop}%
		\bibitem [{\citenamefont {van Kampen}(2007)}]{Kampen}%
		\BibitemOpen
		\bibfield  {author} {\bibinfo {author} {\bibfnamefont {N.~G.}\ \bibnamefont
				{van Kampen}},\ }\href@noop {} {\emph {\bibinfo {title} {Stochastic Processes
					in Physics and Chemestry}}}\ (\bibinfo  {publisher} {Elsevier},\ \bibinfo
		{year} {2007})\BibitemShut {NoStop}%
		\bibitem [{\citenamefont {Risken}(1989)}]{Risken}%
		\BibitemOpen
		\bibfield  {author} {\bibinfo {author} {\bibfnamefont {H.}~\bibnamefont
				{Risken}},\ }\href@noop {} {\emph {\bibinfo {title} {The Fokker-Planck
					equation}}}\ (\bibinfo  {publisher} {Springer},\ \bibinfo {year}
		{1989})\BibitemShut {NoStop}%
		\bibitem [{\citenamefont {Chin}\ \emph {et~al.}(2010)\citenamefont {Chin},
			\citenamefont {Grimm}, \citenamefont {Julienne},\ and\ \citenamefont
			{Tiesinga}}]{chin2010feshbach}%
		\BibitemOpen
		\bibfield  {author} {\bibinfo {author} {\bibfnamefont {C.}~\bibnamefont
				{Chin}}, \bibinfo {author} {\bibfnamefont {R.}~\bibnamefont {Grimm}},
			\bibinfo {author} {\bibfnamefont {P.}~\bibnamefont {Julienne}}, \ and\
			\bibinfo {author} {\bibfnamefont {E.}~\bibnamefont {Tiesinga}},\ }\href
		{\doibase 10.1103/RevModPhys.82.1225} {\bibfield  {journal} {\bibinfo
				{journal} {Rev. Mod. Phys.}\ }\textbf {\bibinfo {volume} {82}},\ \bibinfo
			{pages} {1225} (\bibinfo {year} {2010})}\BibitemShut {NoStop}%
	\end{thebibliography}
	%\bibliographystyle{apsrev4-1}

\end{document}